# IceCube: A Cubic Kilometer Radiation Detector

Spencer R. Klein, for the IceCube Collaboration*

*Abstract*— **IceCube is a 1 km³ neutrino detector now being built at the Amundsen-Scott South Pole Station. It consists of 4800 Digital Optical Modules (DOMs) which detect Cherenkov radiation from the charged particles produced in neutrino interactions. IceCube will observe astrophysical neutrinos with energies above about 100 GeV. IceCube will be able to separate $\nu_\mu$, $\nu_e$ and $\nu_\tau$ interactions because of their different topologies. IceCube construction is currently 50% complete.**

## I. INTRODUCTION

IceCube, shown in Fig. 1, is a 1 km³ neutrino detector being built to record the interactions produced by astrophysical neutrinos with energies above about 100 GeV [1]. IceCube will observe the Cherenkov radiation from charged particles produced in neutrino interactions, using 4800 optical sensors attached to 80 vertical strings which are deployed in a hexagonal array.

IceCube shares many characteristics with the smaller, laboratory-scale detectors discussed at SORMA. It is a large, segmented tracking calorimeter that measures the energy deposition in segmented volumes of Antarctic ice. It can differentiate between the topologies for electron, muon and tau neutrino interactions. It also has very good timing resolution, which is used to accurately reconstruct muon trajectories and to find the vertices of contained events. The size of IceCube is well matched to the energy scale; a muon with an energy of about 200 GeV travels about 1 km in ice.

Manuscript received June xx, 2008. (Write the date on which you submitted your paper for review.) This work was supported in part by the U.S. National Science Foundation under grant number 0653266, by the NSF Office of Polar Programs and the Physics Division, by the Department of Energy under Contract No. DE-AC02-05CH11231, by the National Energy Research Supercomputing Center, and the University of Wisconsin Alumni Research Foundation.
S. R. Klein is with the Nuclear Science Division, Lawrence Berkeley National Laboratory, Berkeley, CA, 94720, USA ( email: srklein@lbl.gov). and with the Physics Department, University of California, Berkeley. The members of the IceCube collaboration are listed at http://www.icecube.wisc.edu/collaboration/authorlists/2008/4.html.

## II. COSMIC RAYS

A major reason to build IceCube is to find the sources of high-energy cosmic rays [2-4]. Cosmic-rays were first observed almost 100 years ago by Victor Hess. Over the past decades, many experiments have observed the cosmic-ray energy spectrum and composition, from GeV energies up to $3\times10^{20}$ eV. The flux drops rapidly with energy, reaching 1/km²/century at the highest energies. Cosmic-rays have a mixed composition containing mostly nuclei from proton to iron, with at most a small fraction of heavier nuclei and photons.

Despite the decades of effort, we still know very little about the origin of cosmic-rays. At energies up to $10^{15}$ eV, cosmic rays are strongly bent in galactic magnetic fields. They likely originate in our galaxy. Supernovae remnants are the most likely sources. Their strong magnetic fields and shock waves can accelerate charged particles.

Galactic magnetic fields are too weak to confine more energetic particles, which are thought to be primarily extra-galactic. Possible sources are active galactic nuclei (AGNs, galaxies with central supermassive black holes) which emit jets of relativistic particles along their axes. Or, cosmic-rays might be accelerated by the sources of gamma-ray bursts (GRBs). GRBs are believed to originate in the collapse of supermassive stars and/or mergers of black holes and/or neutron stars. Either of these sources may provide appropriate conditions to accelerate nuclei to ultra-relativistic energies.

The most energetic cosmic rays have limited ranges. At energies above about $4\times10^{19}$ eV, cosmic protons are excited by collisions with the $3^0$K microwave background radiation, creating a $\Delta$ resonance. The decaying $\Delta$ emits a lower-energy proton. This energy loss limits the range of more energetic protons to about 100 Megaparsecs [5]. Heavier nuclei are photodissociated by interactions with the microwave background; this leads to a similar range limitation.

Further, all but the most energetic cosmic-rays are bent in the intergalactic magnetic fields and so do not point back toward their origins. At energies above $6\times10^{19}$ eV, bending by interstellar magnetic fields may be tolerable. The Auger collaboration has found evidence that some cosmic-rays may point toward nearby (within 75 Megaparsecs) AGNs [6]. However, the Fly's Eye collaboration does not observe this correlation [7].

In the absence of definitive correlations, we must consider other messengers. TeV photons have been observed from some nearby sources, such as supernovae and some nearby

AGNs. At energies above a few TeV, photons interact with interstellar photons, forming $e^+e^-$ pairs; like protons and heavier nuclei, these photons also have a limited range.

In contrast, neutrinos have very small cross-sections and so can freely travel cosmic distances. They are the only particle able to probe high-energy accelerators out to cosmic distances. Here, we focus on the neutrinos with energies above about 100 GeV which are most relevant for understanding cosmic-ray acceleration. These neutrinos are produced in $\pi^\pm$ decays, $\pi^\pm \longrightarrow \mu^\pm \nu_\mu$, followed by $\mu^\pm \longrightarrow e^\pm \nu_\mu \nu_e$, producing a 2:1 ratio of $\nu_\mu$:$\nu_e$. IceCube cannot differentiate between $\nu$ and anti-$\nu$, so we will combine the two particles. Over long distances, neutrino oscillations change this 2:1 ratio into a 1:1:1 ratio of $\nu_\tau$:$\nu_\mu$:$\nu_e$. The charged pions are produced in incidental 'beam-gas' interactions, whereby the nucleons under acceleration interact with either gas or photons present in the accelerator. If cosmic-rays are heavier nuclei, $\nu_e$ may also be produced by nuclear beta decay of unstable isotopes produced in spallation.

The neutrino flux from cosmic-ray accelerators has been estimated by two methods. The first uses the measured cosmic-ray flux and the estimated photon and matter densities at acceleration sites. The second extrapolates the measured TeV photon flux to higher energies, assuming that the photons are from $\pi^0$ decay. That leads to an estimate of the number of $\pi^\pm$. Both approaches find similar neutrino fluxes, and both lead to a similar conclusion: that a neutrino detector with an area of $\sim$ 1 km$^3$ is needed to observe neutrinos from astrophysical sources.

### III. EARLY LARGE NEUTRINO DETECTORS

For obvious cost reasons, a 1 km$^3$ detector must use a natural detecting medium. One approach to such a large detector is to search for optical Cherenkov radiation from charged particles produced in neutrino interactions. Three media have been proposed: seawater, freshwater (in a lake), and Antarctic ice. All three have advantages and disadvantages. Water has a very long scattering length but relatively short absorption length. Seawater has high backgrounds from $^{40}$K decays and bioluminesence, while the available freshwater lakes suffer from limited size. On the other hand, in ice, the scattering length is shorter than in water, and, once deployed, detector hardware is not recoverable. All three approaches have been pursued. The DUMAND collaboration proposed a large seawater detector back in the 1980's. Currently, the ANTARES, NESTOR and NEMO collaborations are working on detectors in the Mediterranean Sea. A Russian-German collaboration has built a detector in Lake Baikal [8].

Neutrino detection in ice was pioneered by the AMANDA collaboration. It requires a thick ice sheet, so AMANDA was built at the Amundsen-Scott South Pole station, where the ice is about 2800 m deep. The collaboration drilled holes in the ice using a hot water drill, and lowered strings of optical sensors before the water in the hole refroze.

AMANDA deployed its first string on Christmas Eve 1993, at a depth of 800-1,000 m. It was quickly found that the ice had a very short scattering length, less than 50 cm. This was explained by small (50 μm) air bubbles in the ice. Fortunately, at the higher pressures present at ice depths greater than 1400 m, these bubbles collapse. With this understood, in 1995-6 AMANDA deployed 4 strings with detectors mounted between 1500 and 2000 m deep. These detectors worked as expected, and AMANDA detected its first neutrinos [9]. This success led to AMANDA-II, which, by 2000 consisted of 19 detector strings holding 677 optical sensors. Since 2000, AMANDA-II has been recording about 1,000 neutrino events per year.

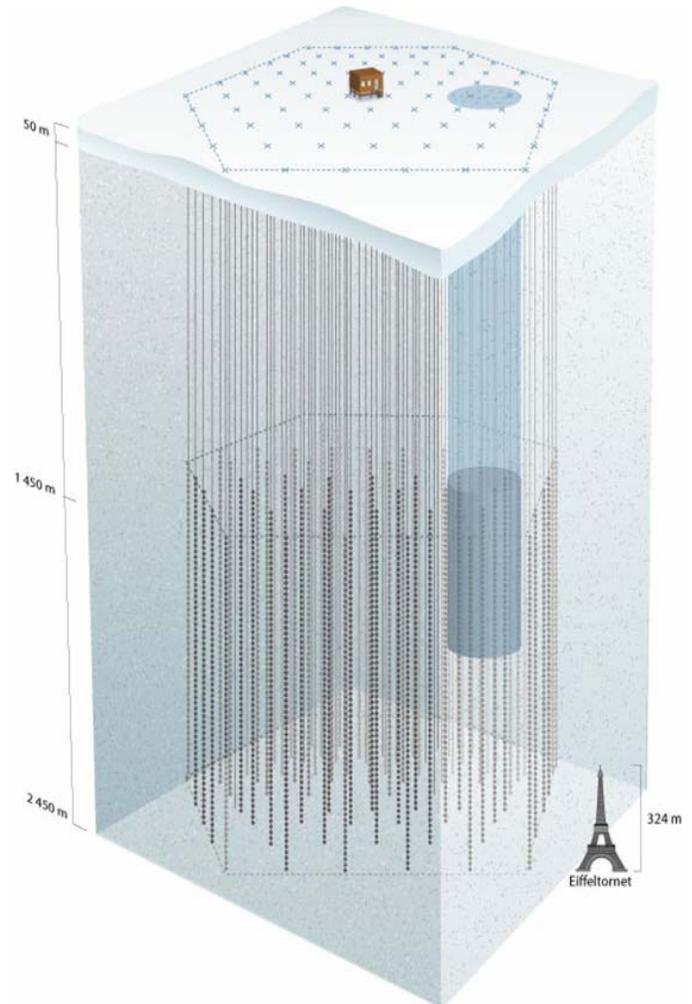

Fig. 1. Schematic of the IceCube detector, showing the 80 strings. The dark cylinder shows the volume of AMANDA.

However, despite this success, the limitations of AMANDA were becoming obvious. It was too small, and the technology did not lend itself to easy expansion. The AMANDA optical sensors consisted of photomultipliers with resistive bases in a pressure vessel. High voltage was generated on the surface, and analog signals were returned to the surface. Since AMANDA was a prototype detector, several transmission media were tried: coaxial cables, twisted pairs, and, later, optical fibers. The 2.5 km long coaxial cables and twisted pairs dispersed the PMT pulses, with single photoelectron pulses broadened to $\sim$ μs widths, while the analog optical fibers had a very limited dynamic range. Further, the system was finicky, and not all of the optical modules survived the high pressures present when the water in the drill holes froze. Finally, AMANDA consumed considerable electrical power and required yearly, manpower-intensive calibrations. IceCube was designed to avoid these problems.

## IV. ICECUBE HARDWARE

IceCube was designed to be much simpler to deploy, operate and calibrate than its AMANDA predecessor. When it is complete in 2011, it will consist of 80 strings of photomultipliers, each containing 60 digital optical modules (DOMs). The strings are placed on a 125 m hexagonal grid. DOMs are placed on a string with 17 m spacing, between 1450 m and 2450 m below the surface. The surface electronics are in a counting house located in the center of the array.

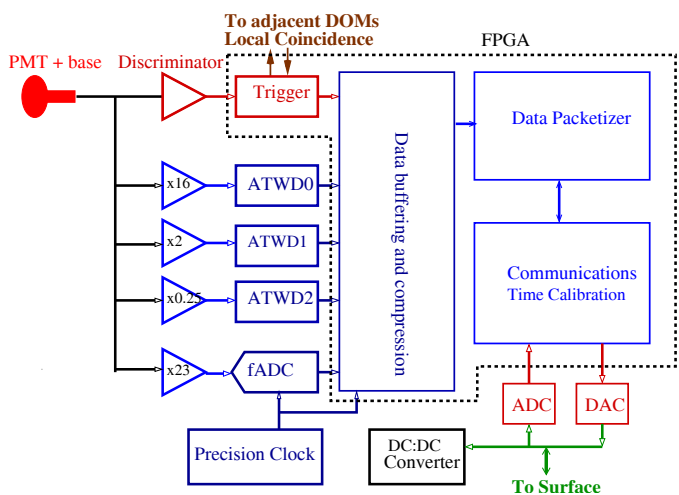

Fig. 2. A block diagram of the IceCube main board electronics.

Each string of 60 DOMs is supported by a cable that contains 30 twisted pairs (each pair is connected to two DOMs in parallel), plus a strength member and a protective covering.

In addition to the deeply buried DOMs, the IceCube Observatory includes a surface air shower array known as IceTop. IceTop will measure the cosmic-ray energy spectrum and composition, above a threshold energy of about 300 TeV. Combined measurements of electromagnetic showers (on the surface) and deep underground muons with IceCube provide a useful technique for measuring the nuclear composition of cosmic-rays. Combined events can also be used to check the pointing accuracy of IceCube. IceTop consists of 160 ice filled tanks, each 1.86 m in diameter. There are two tanks near the top of each string. Two DOMs are mounted in each tank to detect the Cherenkov photons from charged particles in the air shower.

The main task in IceCube construction is drilling holes for the strings of DOMs. This is done with a 5 MW hot-water drill, which generates a stream of 200 gallons/minute of $88^0$C water. This water is propelled through a 1.8 cm diameter nozzle at a pressure of 200 pounds/square inch, melting a hole through the ice. Drilling a 2500 m deep, 60 cm diameter hole takes about 40 hours. Deploying a string of DOMs takes about another 12 hours.

Because of the Antarctic weather, the high altitude and the remote location of the South Pole, logistics is a key issue for IceCube. The construction season lasts from roughly November through mid-February. Everything needed must be flown to the Pole on ski-equipped LC-130 transports planes.

IceCube construction began in 2004/5, when the first string was deployed. In 2005/6, eight additional strings were deployed, and, during 2006, data was taken with nine strings. In 2006/7, thirteen strings were deployed, followed by eighteen in 2007/8, leaving the detector half done.

## V. DOM HARDWARE

Each DOM contains a downward-facing 10" (25 cm) Hamamatsu R7081-02 photomultiplier tube and associated electronics in a 35 cm diameter pressure sphere. The PMT has a standard bialkali photocathode (Sb-Rb-Cs, Sb-K-Cs), with a peak quantum efficiency of about 25% at 390 nm. The minimum useful wavelength of about 350 nm is set by absorption in the pressure sphere. The electronics includes a Cockroft-Walton high voltage power supply, electronic timing calibration systems, light emitting diodes for photonic calibrations, and a complete data acquisition (DAQ) system. The PMTs are currently run at a gain of $10^7$, with typical high voltages of 1300-1500 volts. An average single photoelectron produces a pulse about 10 mV in amplitude and 5 nsec width into the 43 Ω impedance of the DAQ system. The charge resolution for single photoelectrons is about 30%. The DAQ system is designed to record the arrival time of all detected photoelectrons, with a relative precision of better than 5 nsec, across the entire array.

A block diagram of the DAQ system is shown in Fig. 2. The central elements of the DAQ hardware are two waveform digitization systems, the Analog Transient Waveform Digitizer (ATWD) and the fADC ('fast' ADC). A digitization cycle is initiated by a discriminator trigger; the threshold is set at a voltage corresponding to about 1/4 photoelectron. When this happens, the FPGA will start ATWD and fADC digitization on the next clock edge. To make up for delay in the trigger circuit, the signal goes through a 75 nsec delay line before the digitizers. This delay line limits the system bandwidth to about 100 MHz.

The ATWD digitizer uses a custom switched-capacitor array chip. Each ATWD chip includes four parallel inputs, each with 128 capacitors. When launched, the system acquires data at 200 to 900 megasamples per second (MSPS); IceCube runs the ATWDs at 300 MSPS, providing 400 nsec of recording capacity. Three ATWD channels are run in parallel, with input gains in the ratio of 16:2:1/4, providing more than 14 bits of dynamic range. After acquisition, the voltages on the capacitors are digitized with 128 10-bit Wilkinson ADCs, each multiplexed to the four capacitors which acquire a single time sample. A fourth ATWD input (not shown) is used for electronics calibrations. Each DOM contains two ATWD chips. They are operated in a ping-pong fashion – while one is digitizing, the other is live; this greatly reduces the dead time. The fADC digitizer uses a 10-bit, 40 MSPS commercial ADC chip. When triggered, the system records 256 samples, covering 6.4 μs.

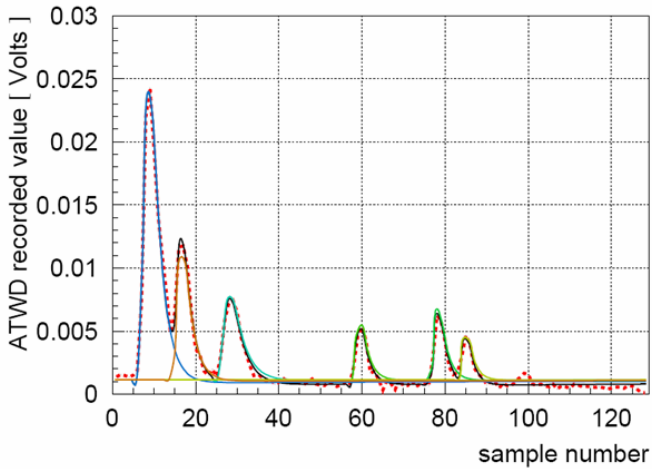

Fig. 3. The ATWD digitizer output from a typical event. Multiple pulses are shown. The waveform is decomposed into a list of photon arrival times, which is used for event reconstruction.

Each DOM also contains a 'flasher' board, which has 12 blue (405 nm) LEDs mounted around its edges. These LEDs are used for a variety of calibrations, measuring light transmission and timing between different DOMs, to check the DOM-to-DOM relative timing and study the optical properties of the ice.

The entire system is controlled by a 400k gate Altera Excalibur FPGA, which incorporates an ARM9 hard-core CPU. The FPGA controls the data acquisition and digitization cycle, compresses (losslessly) and formats data for transmission to the surface, and oversees calibrations.

Data is transmitted to the surface via a single twisted pair, which also provides ±48 VDC (96 volts total) power. Each DOM consumes about 3.5 W. The cable also includes local coincidence circuitry, whereby DOMs communicate with their nearest neighbors; they can also pass messages onward. A more robust connector is used than in AMANDA, and a higher fraction of IceCube OMs survive 'freeze-in.' On the surface, the cables are connected to a custom PCI card in a PC; the remainder of the system is off-the-shelf.

IceCube DOMs have several operating modes. They are currently operating in "Hard Local Coincidence" mode: data is only saved when a neighbor (either nearest or next-to-nearest) DOM also sees a signal within 1 μs. In "Soft Local Coincidence" mode, an abbreviated 'coarse charge stamp' is saved even for isolated hits. It consists of the largest 3 fADC samples out of the first 16 samples. Saved data is formed into packets for transmission to the surface.

The system uses a 40 MHz system clock. Since this clock is used to record the hit times, a precision oscillator is used. The oscillator has frequency stability (Allen variance) of better than $\delta f/f < 10^{-10}$. Despite this accuracy, maintaining the required 5 nsec precision requires frequent synchronization.

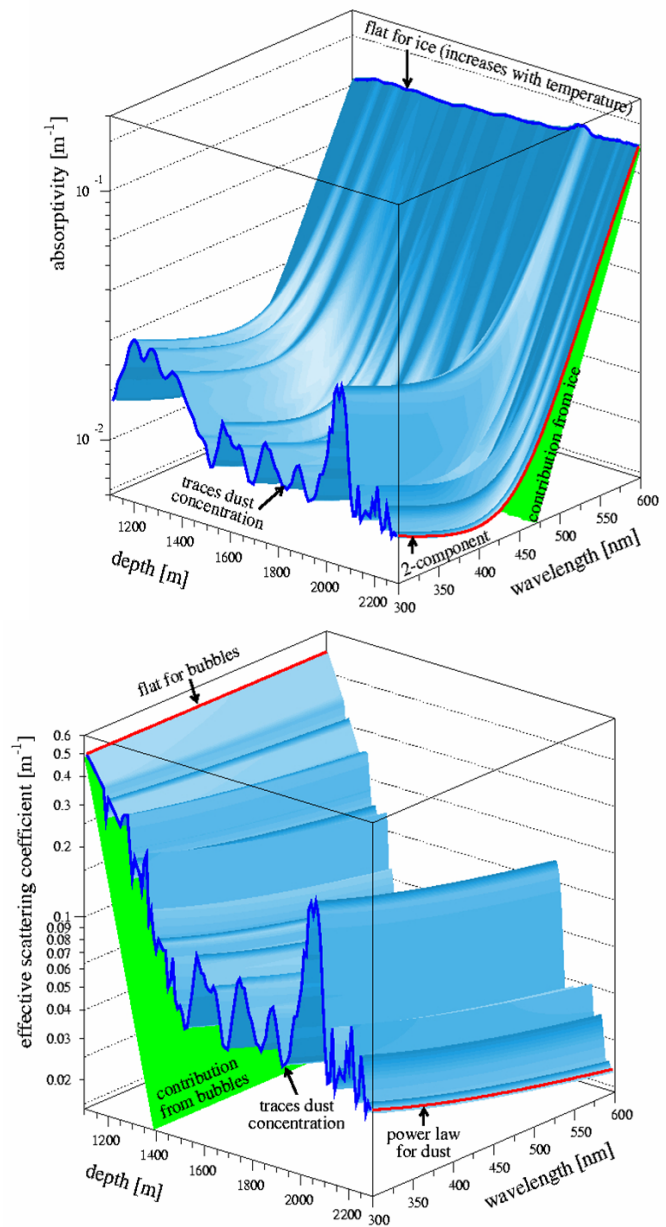

Fig. 4 Absorption (top) and scattering (bottom) lengths of light in South Polar ice, as a function of depth and wavelength. From Ref. [11].

Timing calibrations are performed automatically every few seconds (currently once every 0.5 s). During each calibration, the surface electronics sends a timing signal down to each DOM, which waits a few μs until cable reflections die out, and then sends an identical signal to the surface. To maintain the symmetry, both the surface and DOM electronics use identical DACs and ADCs to send and receive signals. With the symmetric setup, transmission times in the two directions are identical. Even though the 3.5 km cable transmission widens the signals to ~1 μs, the transmission time is determined to better than 3 nsec [10]. This accuracy is maintained across the entire array; it has been verified using muons and artificial light sources. The software tracks the timing difference between the in-DOM oscillators and a surface based master clock, and appropriate corrections are applied to the data.

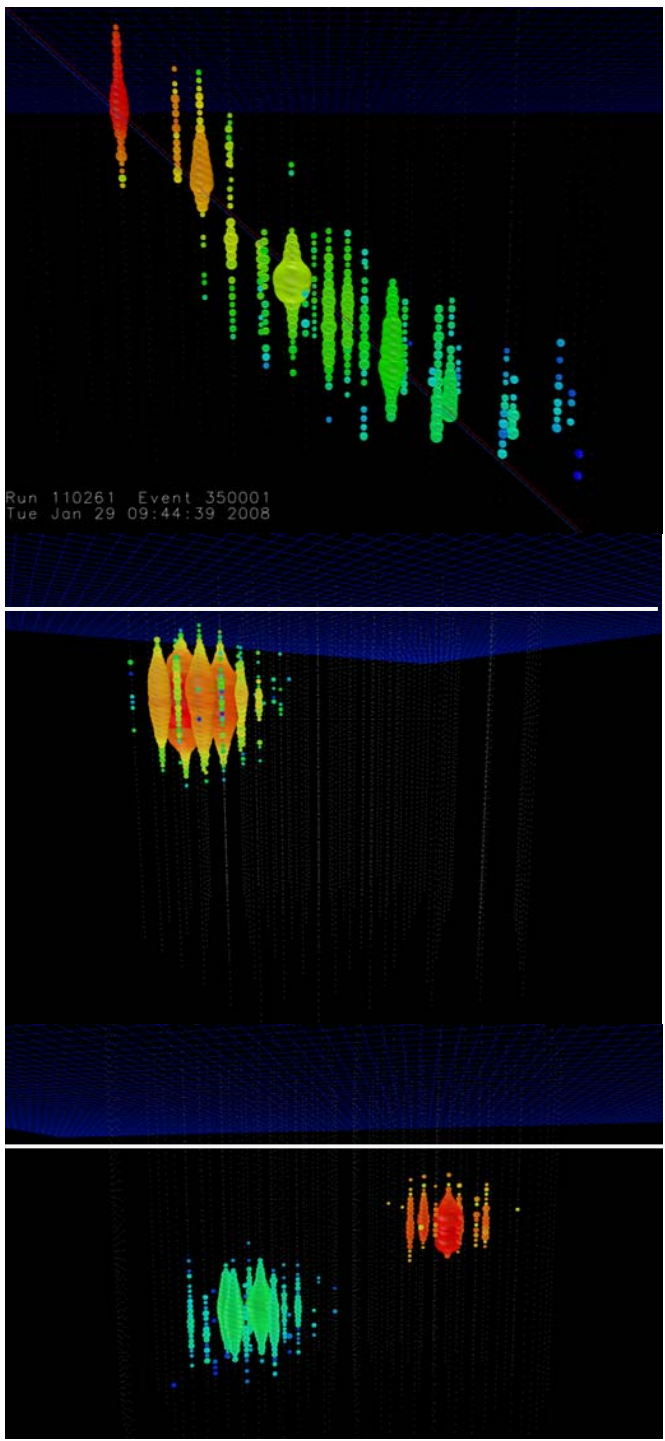

Fig. 5. IceCube event displays for (top) a muon or muon bundle (multiple muons) in IceCube 40 (the 40 string configuration running in 2008), a simulated $\nu_e$ (middle) and a simulated $\nu_\tau$ (bottom). The latter shows the classical 'double-bang' topology. Each dot is from a single struck DOM. The size of the circles indicates the number of detected photons, while the color gives the time, from red (earliest) to blue (latest). The 125 m string spacing and the 17 DOM-to-DOM spacing give the scale of the events. From Ref. [3].

Amplitude calibrations are done using an ultraviolet (peaked around 374 nm) LED that is mounted on the main electronics board. It is flashed repeatedly at low intensity (<< 1 photoelectron in the PMT). A charge histogram is accumulated in the FPGA and sent to the surface, where it is fit to find the single photoelectron peak. This is done for a range of high voltages, and the system is then set to the correct HV to give $10^7$ PMT gain. These calibrations are extremely stable.

Amplitude linearity calibrations take advantage of the 12 LEDs on the calibration board. The LEDs are flashed individually, and then together, providing a ladder of light amplitudes that can be used to map out the saturation curve.

One other critical requirement for the IceCube hardware is high reliability without maintenance. Once deployed, it is impossible to repair a DOM, so the system was designed for very high reliability. About 98% of the DOMs survive deployment and freeze-in completely; another 1% are impaired, but usable (usually, they have lost their local coincidence connections). Post-freeze-in, reliability has been excellent, and the estimated 15-year survival probability is 94%.

## VI. THE ICE IN ICECUBE

The ice surrounding the DOMs is a critical part of IceCube. Both absorption and scattering are significant. Both are strongly affected by impurities in the ice. These impurities are a reflection of the impurities in the air when the ice was first laid down as snow. This happened over roughly the last 100,000 years. Because of variations in the long-term dust level in the atmosphere during this period, as well as the occasional volcanic eruption, the impurity concentrations are depth dependent.

Much effort has gone into measuring the optical properties of the ice, using artificial light sources and in-situ measurements. In AMANDA and IceCube, studies have been done using LEDs and lasers that emit at a variety of wavelengths. By measuring the arrival time distributions of photons at different distances from a light source, it is possible to measure both the attenuation length and scattering length of the light. These measurements, although useful, suffer from a limited resolution in depth [11].

Higher resolution depth-dependence measurements of the ice properties come from a 'dust logger' which is lowered down a water-filled hole immediately after drilling. The dust logger shines a thin beam of 404 nm light into the ice, and measures the reflected light [12]. This provides a measure of the ice properties on a depth scale given by the width of the emitted beam – a few mm.

Figure 4 shows our understanding of ice absorption and scattering distances, as a function of depth and wavelength. At depths below about 1400 m, air bubbles are present in the ice. These bubbles greatly limit the optical scattering length in the ice. At deeper depths, the broad peaks in both the absorption and scattering lengths are due to dust in the ice. Not visible are the very narrow peaks due to thin layers of dust produced by volcanoes. The underlying scattering lengths are derivable from Mie scattering. The exponential rise in absorption at long wavelengths is believed to be due to molecular absorption.

## VII. DATA TRIGGERING AND FILTERING

Data collected by the DOMs and sent to the surface is time-sorted, combined into a single stream, and then monitored by a software trigger. IceCube uses two trigger criteria and may add a third. The main trigger is based on multiplicity; it selects time intervals where eight DOMs (with local coincidences) fired within 5 μs. This collects most of the neutrino events. In 2008, a string trigger was added; it selects time intervals when five out of seven adjacent DOMs fired within 1.5 μs. This trigger has improved sensitivity for low energy (down to 100 GeV) events, especially upward going muons. A third, 'topological' trigger is also under consideration; it will be optimized for low-energy horizontal muons. When any trigger occurs, all data within the ±10 μs trigger window is saved, becoming an event. If multiple trigger windows overlap, then all of the data from the ORed time intervals are saved as a single event.

The total trigger rate (for 40 strings) is about 1,400 Hz. The majority of the triggers (about 1,000 Hz) are due to cosmic-ray muons, with the rest divided among other sources, including IceTop.

Triggered data is reconstructed by an on-line filter system and selected events are transmitted via satellite to the Northern hemisphere. The filters use simple criteria, 'first-guess' reconstruction algorithms and simplified maximum likelihood fitting. Current filters select upward going muons, cascades ($\nu_e$, $\nu_\tau$ and all-flavor neutral current interactions), extremely high energy events, starting and stopping events, and air showers seen in IceTop. Currently, about 6% of the events are selected by these filters, comprising about 32.5 Gbytes/day. The remainder of the data is stored on tapes at the South Pole station. The tapes are carried north during the austral summer.

## VIII. EVENT RECONSTRUCTION

In the Northern hemisphere, events are reconstructed using maximum-likelihood fitting techniques. Events are fitted to templates representing different decay modes. Figure 5 shows examples of three different interaction topologies in IceCube [2].

Figure 5 (top) shows an actual (data) muon or muon bundle (group of parallel muons from an air shower). The tracks are visible over more than 1 km. This long lever arm allows for good directional reconstruction, better than 1 degree. Of course, for shorter tracks, the resolution degrades. It is also possible to estimate muon energies by either the length of their tracks, or by measuring the specific energy loss; at energies above 1 TeV, muon energy loss (dE/dx) is proportional to the muon energy.

Figure 5 (middle) shows a simulated $\nu_e$ interaction which produces a compact deposition of energy; this is known as a 'cascade.' Cascades are also produced by neutral current neutrino interactions and low-energy (below 1 PeV) $\nu_\tau$ interactions. Although there is very little directional information, cascade energies may be determined to within a factor of 2.

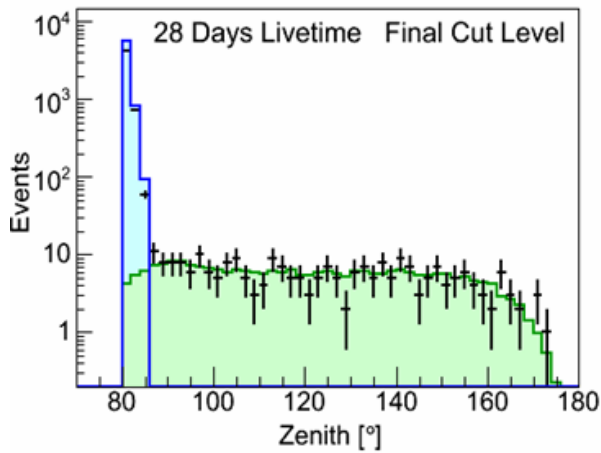

Fig. 6. The azimuthal angle for downward-going, or near downward-going muons in IceCube 40, after tight cuts, compared with the results of cosmic ray muons (blue) and neutrinos (green) simulations. The coincident muon background is largely eliminated (4 downward going events expected) and not shown here.

Figure 5 (bottom) shows a simulated few-PeV $\nu_\tau$ interaction forming a classic 'double-bang' topology. The interaction produces one cascade when the $\nu_\tau$ interacts. That interaction produces a τ, which, at PeV energies, can travel hundreds of meters before decaying. The second cascade comes when the τ decays. Several other τ decay modes are under study in IceCube.

Other topologies are also being studied. For example, a $\nu_\mu$ interacting in the detector will produce a hadronic shower from the struck nucleus, in addition to the μ track. Muons can also stop in the detector.

Of course, the most common events are downward going muons produced in cosmic-ray air showers. In triggered events, cosmic-ray muons outnumber neutrino induced muons by about 500,000:1. Rejection of this background is a significant difficulty which must be dealt with in event reconstruction.

Events are reconstructed by fitting them to one of these hypotheses. The starting points for these fits are the results of 'first guess' methods. For muons, the main first guess method fits a moving plane to the launch times in the DOMs [13]. For a muon, the plane should have a velocity near the speed of light. An alternate approach uses the measured charge deposition to the 'long axis' in events such as in Fig. 5 (top).

The maximum likelihood fitter finds the likelihood for different track positions and directions, and, optionally, energy. To do this, it uses functions which model the light propagation, giving the probability distribution for a photon radiated from a track with a given orientation to reach a DOM at a given perpendicular distance and orientation as a function of time. These functions are precalculated using a simulation that tracks photons through the ice, and stored in a 7-dimensional histogram [14]. One of the dimensions is depth, incorporating the depth dependence of the optical properties of the ice.

Because of the high rate of downward going muons, it is not enough to select events with the most likely reconstruction as upward going [15]. Fairly stringent cuts must be applied to eliminate tracks with reasonable likelihoods for being downward going. This can be done by cutting on the estimated errors from the likelihood fit, which can act as a stand-in for the depth of the minimum in the likelihood function. Alternately, one can perform a Bayesian reconstruction, weighting fits to different zenith angles by the relative size of the signal in that direction (effectively requiring that the upward going hypothesis be much more likely). The exact cuts are analysis-dependent, since different analyses are interested in signals from different energy ranges and zenith angles.

IceCube is big enough that there is also a significant background due to random coincident muon events, whereby two (or more) muons from independent cosmic-ray air showers traverse the detector in the course of one event. Specific algorithms have been developed to find and reject these events, by separating hits from the two tracks based on their separation in space and/or time.

After these cuts, a relatively clean sample of well-reconstructed neutrino events remains, as is shown in Fig. 6. There remains an irreducible background of atmospheric neutrinos produced by cosmic-ray air showers in the northern hemisphere. In 1 year (about 320 live days) of IC40 data, we expect about 5,000 atmospheric $\nu_\mu$ interactions. The atmospheric $\nu_e$ background is about two orders of magnitude lower and the atmospheric $\nu_\tau$ background is almost absent.

The lower backgrounds make the two latter channels attractive avenues to search for extraterrestrial neutrinos. In searches for point sources of neutrinos, off-source regions are used to directly measure the background level [16]. Diffuse neutrino analyses use the fact that the energy spectrum of the atmospheric neutrinos is much softer than for extra-terrestrial neutrinos; by selecting high energy events, one can largely remove the atmospheric background [17]. Current diffuse searches have most of their sensitivity above 100 TeV.

## IX. FUTURE PLANS

IceCube completion is scheduled for 2011. In addition to the 80 baseline strings, we are also developing a Deep Core infill array. Deep Core will consist of 6 additional strings with a smaller, 72 m grid spacing. The DOMs will use new phototubes with 25% higher quantum efficiency. They will be spaced every 7 m in the deepest, clearest 350 m of ice. In addition, the rest of IceCube will serve as a veto region surrounding Deep Core, allowing for the rejection of cosmic-ray muons and other non-contained backgrounds. The higher granularity, improved optical sensitivity and surrounding veto will give Deep Core a much lower threshold than IceCube, as low as 10 GeV.

IceCube collaborators are also studying prototype radio and acoustic neutrino detectors. These are sensitive to coherent radio-Cherenkov emission from neutrino-induced electromagnetic and hadronic showers and the shock-wave produced by local heating from neutrino induced showers respectively. The radio and acoustic signals should have much larger absorption lengths than light, so these techniques might be usable to build a much larger (100 km$^3$) array than IceCube. However, because of the technique used, the array would have a much higher energy threshold, perhaps $10^{17}$ eV.

## X. CONCLUSIONS

The 1 km$^3$ IceCube neutrino observatory detects Cherenkov radiation from charged particles produced in neutrino interactions. With its 4800 digital optical modules, IceCube acts like a tracking calorimeter, recording the pattern of energy deposition in the ice. Each DOM includes a complete data acquisition system. IceCube construction is 50% complete and the system is working well with very high reliability.

The segmentation gives IceCube the capacity to separate the different topologies from $\nu_\mu$, $\nu_e$ and $\nu_\tau$ interactions. We have developed reconstruction methods that effectively separate upward going muons from $\nu_\mu$ interactions from the much more intense cosmic ray muon background. These methods achieve an angular resolution of better than 1 degree for long tracks.